# Compact objects in globular clusters

Globular clusters and compact objects (i.e. black holes, neutron stars, and white dwarfs) are intimately intertwined. The dense stellar environments in globular clusters lead to high rates of production of interesting classes of compact objects, such as accreting compact objects, binary pulsars, and, perhaps even intermediate mass black holes. At the same time, the presence of compact objects in globular clusters may affect cluster evolution. The importance of compact objects in globular clusters, and some of the possibilities they represented, were first appreciated in the mid-1970's. However, the observational capabilities, in both X-ray and optical wavelengths, were not available to explore these points more carefully until the last decade or so, with the launches of the Chandra and X-ray Multiple Mirror (XMM)-Newton observatories in X-rays, and the Hubble Space Telescope in optical. A similar boom in radio pulsar searches in globular clusters has taken place in the last few years due to the Green Bank Telescope. In this paper, we will review the historical and modern work on the populations of compact objects in globular clusters.

**The first discoveries**
The launch of *Uhuru* allowed, for the first time, a relatively high angular resolution survey of the X-ray sky. One of its most surprising results was the discovery that many bright X-ray sources are located in the Milky Way's globular clusters (Clark 1975; Clark, Markert & Li 1975). Around this same time, it was becoming clear that many neutron stars were born with large space velocities – space velocities much larger, in fact, than the escape velocities of globular clusters (Crampton 1974; Manchester, Taylor & Van 1974). It was thus suggested that there needed to be multiple modes of stellar collapse – one which produced the fast radio pulsars and young accreting objects far from the plane of the Galaxy, and one which produced objects which could be retained within globular clusters (Katz 1975).

It had already been suggested that globular clusters might be remnants of quasars, and that they hence might contain rather massive black holes (Wyller 1970), perhaps even containing binary black holes which would have been detectable with even the rather primitive gravitational wave detections of the 1970s. However, this suggestion never attained much prominence, probably because it was already known that the nuclei of nearby galaxies such as M87 were lower luminosity versions of the same phenomenology that produced quasars, implying that the quasar phenomenon happens at the centers of galaxies rather than on the outskirts.

After the discovery of X-ray emission from globular clusters, the suggestion that they might harbour massive black holes was raised again, by Bahcall & Ostriker (1975). They argued that massive clusters would retain gas from the planetary nebula ejections that happen within them, so that there would be some gas for the black hole to accrete. Unfortunately, one of their clearest predictions, that these massive black holes should not show strong variability was quickly ruled out.

**Black holes or neutron stars?**

Shortly after the discoveries of X-ray sources in globular clusters, another important discovery was

made – that these cluster sources showed strong pulses of X-ray emission – Type I X-ray bursts. In fact, the first Type I X-ray bursts were seen from the globular cluster NGC 6624 (Grindlay et al. 1976). For a while, these bursts were considered to possibly be related to accretion onto black holes of a few hundred solar masses in globular clusters (Grindlay & Gurksy 1976), but it was quickly found that the spectra of the bursts were in good agreement with what would be expected from blackbody radiation from a source with the size of a neutron star (Swank et al. 1977). As the Type I bursts are believed to be due to thermonuclear burning (Woosley & Taam 1976) on the surface of a neutron star, the detection of bursts from all the X-ray bright globular clusters (see Liu, van Paradijs & van den Heuvel 2001 and references within) led to the interpretation that these objects are all accreting neutron stars.

The case in reality is not so simple, as it is possible for a single globular cluster to have multiple bright X-ray sources. In fact, with the Chandra observatory's high spatial resolution, it was recently proved that M15 has two bright X-ray sources – the well-known AC 211, an edge-on X-ray source with a relatively bright radio counterpart and an optical counterpart which shows a 17 hour orbital period, and the more recently discovered M15 X-2, which is actually the brighter of the two sources, and is an ultracompact X-ray binary, with an orbital period of about 20 minutes (Dieball et al. 2005).

**Pulsars**
In 1982, Ali Alpar and collaborators suggested that low mass X-ray binaries with neutron star accretors evolve into millisecond pulsars, with the accretion process helping to diminish the neutron star's magnetic field, and the angular momentum accreted spinning up the neutron star. Shortly thereafter, it was suggested that, since globular clusters host large numbers of low mass X-ray binaries, they should host similarly large numbers of millisecond pulsars (Katz 1983). Searches for these pulsars began shortly thereafter.

Two papers in 1985 reported a steep spectrum radio source in the core of the globular cluster M28 (Hamilton, Helfand & Becker 1985; Mahoney & Erickson 1985). The source's spectrum was, in fact, so steep that it was already presumed to be a radio pulsar when its pulsations were actually detected by Lyne et al., in 1987. Searches for pulsars in globular clusters have progressed over the past two decades to the point where more than 100 are now known, in 24 different globular clusters (see Paulo Freire's continuously updated web page http://www.naic.edu/~pfreire/GCpsr.html for a full catalog). Two clusters, 47 Tuc and Terzan 5, now show at least 20 pulsars (Camilo et al. 2000; Ransom et al. 2005).

New instrumentation, in the form of the new Green Bank Telescope with its 100m dish, and the Pulsar Spigot (Kaplan et al. 2005), with its large bandwidth and excellent spectral resolution have pushed this work forward in the past few years. With older instrumentation, it was difficult, if not impossible, to detect pulsations from globular clusters behind the plane of the Galaxy. Foreground free electrons dispersed the low frequency radiation from these pulsars, spreading out the pulsations in time, and making the changes in brightness as a function of phase undetectable. The Green Bank Telescope has a much larger effective area at high radio frequencies than previous steerable radio telescopes did. The combination of this large effective area, plus better spectral resolution and more bandwidth allows higher frequency observations to be done with good sensitivity, and also allows for more effective "de-dispersion" of the pulses.

The results from the Green Bank Telescope are only starting to come in. It was long known, for example, that Terzan 5 should have many pulsars, because its integrated radio flux was quite large (Fruchter & Goss 2000). On the other hand, only three pulsars in Ter 5 were known before 2005. Another 31 pulsars have been discovered so far in this clusters in the last two years (Ransom et al. 2005; Hessels et al. 2006). Overall, roughly 50% of all known globular cluster pulsars have been discovered with the Green Bank Telescope, and all of these discoveries have been made since 2003.

**Which clusters have the most X-ray sources?**

The number of X-ray sources found in a globular cluster is expected to correlate with several variables. Since these sources are formed dynamically, the most important is likely to be the interaction rate between compact objects and normal stars. Small number statistics for bright X-ray sources hamper this work in the Milky Way, while the difficulty in measuring the spatial profiles of globular clusters in nearby galaxies prevents existing extragalactic studies from being useful.

The situation is not nearly so dire as it seems. Even with just the Milky Way data, there were some indications that the most massive, most central and most metal rich globular clusters were most likely to contain X-ray binaries (Grindlay 1987), and these results held up when the data from M31 was added to the sample (Bellazzini et al. 1995). However, given the rather strong correlations of different parameters of the Milky Way's globular clusters with one another (particularly metallicity with galactocentric radius – see Djorgovski & Meylan 1994), and the still small number of X-ray sources, it remained uncertain which were the root causes of the correlations and which were the secondary correlations.

The extragalactic data have provided sufficient statistical samples to solve these problems. By performing discriminant analysis on the X-ray sources in NGC 4472, it was shown that the metallicity correlation is statistically significant, and substantially stronger than the marginal anticorrelation with galactocentric radius (Kundu, Maccarone & Zepf 2002). Numerous other galaxies have since shown the same correlation between metallicity and probability that a globular cluster will host an X-ray source (Di Stefano et al. 2003; Jordan et al. 2004; Minnitti et al. 2004; Xu et al. 2005; Posson-Brown et al. 2006; Chies-Santos et al. 2006; Sivakoff et al. 2007; Kundu et al. 2007). Several compilation papers have, not surprisingly, found these same results for large samples of galaxies (Sarazin et al. 2003; Kim et al. 2006).

The origin of this correlation is not clear. While the metal rich clusters are thought to be a bit younger, on the whole, than the metal poor clusters, the lack of evidence for an excess of X-ray binaries in systems where there are clearly young clusters (Kundu et al. 2003 – see below for more discussion) suggests that the very small age spread in typically elliptical galaxies is unlikely to explain the metallicity effect. Instead, two models in which the metallicity directly affects the binary evolution have been proposed. In the first, irradiation induced winds are driven from the atmosphere of the donor star (Maccarone, Kundu & Zepf 2004). These winds are stronger for brighter X-ray sources than weaker ones, and are stronger for metal poor stars than for metal rich stars, because the atmospheres of metal rich stars can dissipate energy through line cooling, which is not as easily done in metal poor stars (Iben, Tutukov & Fedorova 1997). Alternatively, or perhaps additionally, it may be that the smaller convection zones of metal poor stars relative to metal rich stars at a given mass both inhibit tidal captures, by making it difficult for the excess energy to be dissipated, and by preventing magnetic braking from working efficiently, so that the accretion rate is dramatically suppressed compared to a metal rich star's accretion rate, even once a binary system has formed (Ivanova 2006).

In fact, the strongest correlation with X-ray binary hosting rate of extragalactic globular clusters is with the cluster mass, which is not surprising, because in the Milky Way, there is a strong correlation between the cluster mass and cluster core collision rate (Smits et al. 2006). Attempts to make use of the limited spatial profile data available for extragalactic globular clusters have found that the more concentrated clusters seem to have higher rates of hosting X-ray binaries, but the evidence for this is statistically marginal (Kundu et al. 2002; Jordan et al. 2004).

The key problem is that the pixel sizes on the Hubble Space Telescope instruments are comparable to or smaller than the core radii of globular clusters in Virgo Cluster galaxies, which means that to obtain reasonable core radii, one needs very high signal-to-noise data, and some degree of confidence that the King model or other structural model being used to fit the data reliably describes the data. Virgo Cluster elliptical galaxies (or other galaxies at similar distances) are ideal for most aspects of the study of extragalactic X-ray sources, because a single Chandra pointing of about half a day can detect about 100 X-ray sources, 90% of which are truly associated with the target galaxy. To obtain data on so many globular cluster X-ray sources more nearby requires large surveys of spiral galaxies or the anomalous elliptical galaxy NGC 5128 (which hosts the radio source Cen A). This would require very large HST programs, and has not yet been completed.

Extragalactic globular clusters also allow for potential tests of properties that cannot be considered in the Milky Way. In particular, some theoretical work has suggested that the number of bright X-ray sources in a globular cluster should be a strong function of the cluster age – it should peak typically at ages of a few Gyrs, after all the black holes and neutron stars have formed and found their way to the cluster core through mass segregation, but while there are still relatively heavy main sequence stars present which have higher cross-sections for 2- and 3-body interactions (see, for example, Davies & Hanson 1998).

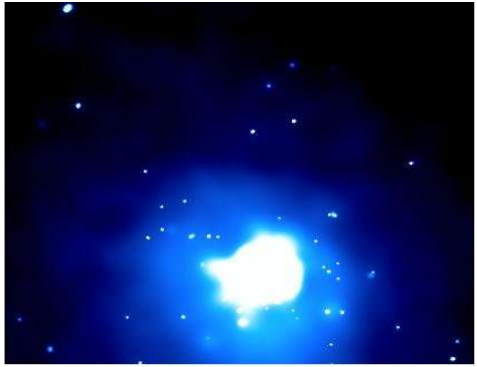

*A smoothed X-ray image of NGC 4472, the brightest galaxy in the Virgo Cluster, from Chandra. The diffuse emission is from the galaxy's reservoir of hot gas, while the point sources are mostly bright X-ray binaries.*

The most reliable method for estimating the ages of extragalactic globular clusters is, at the present time, a comparison of optical colors with infrared colors. The infrared emission is sensitive primarily to light from giant stars, so the infrared colors are dominated by the metallicity of the giant branch. The optical colors are affected partly by metallicity, but also are affected substantially by the temperatures of the turnoff stars in the clusters. Clusters whose optical colors are bluer than expected given their infrared colors are thus likely to be substantially younger than most globular clusters. Unfortunately, to date, the sample of globular clusters which have been observed in the X-rays and have had their ages measured is rather small. The present data do not show any clear affect of age on probability of hosting an X-ray binary (Kundu et al. 2003), but also cannot rule out a factor of 5 or so enhancement in the rate of X-ray binary production in intermediate aged clusters, and so are consistent with some of the parameter space explored by Davies & Hanson (1998).

It is not clear which properties of globular clusters make them most likely to contain millisecond radio pulsars. Certainly the most massive, highest collision rate clusters, such as 47 Tuc and Terzan 5 have large numbers of pulsars, but the selection effects in pulsar searches make it difficult to say more than this. Understanding whether, for example, metallicity affects the number of millisecond pulsars in a globular cluster is a key for understanding why the metallicity effects work for low mass X-ray binaries. If, for example, the scenario for producing the metallicity dependence is one where the evolution of X-ray binaries is accelerated at low metallicity, then there should be no dependence of the number of pulsars on the cluster metallicity. If, on the other hand, the formation rate of X-ray binaries is also affected by the cluster metallicity, then the metal rich clusters should have more pulsars, just as they have more X-ray binaries. Hopefully, with the continuing pulsar searches being made with the Green Bank Telescope, and the reduced selection effects made possible by the use of higher frequencies, we will soon have an answer to this question.

**New thoughts and results on black holes in globular clusters**

There has been considerable debate over time not only about whether globular clusters *do* contain black holes, but whether they *can* contain black holes. The debate started in 1969, when Lyman Spitzer published a paper showing that mass segregation, the process by which the most massive objects in a star cluster sink to the center, would lead to decoupling – essentially the creation of a cluster within a cluster – if the masses of the stars were too different from one another. Thus, in an old stellar population, where the stellar mass black holes, being about 7 or more solar masses, are five times heavier than the neutron stars (the next most heavy objects in the cluster), the black holes should become decoupled. Numerical work quickly followed, showing, as expected from theory, that these decoupled clusters would evaporate very quickly, ejecting most or all of the clusters' black holes (Kulkarni et al. 1993; Sigurdsson & Hernquist 1993; Portegies Zwart & McMillan 2000).

Recently, however, these controversies have been re-kindled. Both theoretical work and observational work has suggested the presence of intermediate mass black holes in globular clusters. Theoretical work has suggested a few ways to avoid the problems of retaining black holes in the centers of globular clusters. Miller & Hamilton (2002) have pointed out that the initial mass functions of metal poor stellar systems may be top-heavy, and that such stars will also have weak stellar winds. This may allow the occasional production of a black hole through normal stellar evolutionary channels of order 50 solar masses, which allows for mergers of black holes to take place without the gravitational radiation rocket effect (Redmount & Rees 1989) ejecting the binary system in its late stages. The Newtonian dynamical effects will also fail to eject such heavy black holes.

An alternative is the runaway merger of stars at the center of a star cluster. Certain young, dense star clusters have core collapse timescales shorter than the main sequence lifetimes of stars. As a result, the stars in these clusters's cores will merge on a timescale fast compared with their main sequence lifetimes. This has the potential to allow the production of a star of about 1000 solar masses in the center of the cluster (Portegies Zwart & McMillan 2002), but there remains considerable debate about whether the stellar winds from such an object would limit the stellar mass to something considerably smaller. Considerably more work needs to be done to determine whether these mergers are a viable mechanism for producing heavy black holes.

Observational work has also suggested that some globular clusters contain intermediate mass black holes, although this work is also speculative at the current time. The chief evidence for these intermediate mass black holes comes from velocity dispersion curves, and, in some cases, evidence of rotation (Gerssen et al 2002; Gebhardt, Rich & Ho 2002, 2005). The velocity dispersion profiles indicate a sharp increase in mass-to-light ratio in the very centers of globular clusters, with the excess dark masses However, it has also been shown that these results can be the consequence of mass segregation in globular clusters (Baumgardt et al. 2003a&b); in globular clusters, the heaviest objects will be the ones with the lowest mass-to-light ratios – neutron stars and white dwarfs. In fact, arguments for (Newell et al. 1976) and against (Illingworth & King 1977) the existence of an intermediate mass black hole in M15 with largely the same substance had been made almost a quarter century earlier, but these papers seem to have been forgotten over time.

Alternative tests, aside from kinematics, are thus important for determining whether there are massive black holes in globular clusters. In principle, gravitational microlensing would be ideal, as it would provide a very clear diagnostic. Unfortunately, this requires a background star to move very close to the location of the black hole. Since the effects of microlensing are maximized for the case where the lens is halfway between the lensed object and the observer, the probability of having a substantial lensing event is quite small for most globular clusters – their background objects would be halo stars.

The alternative is to return to the suggestion of Wyller (1970), and Bahcall & Ostriker (1975), that a massive black hole in a globular cluster should be able find some gas to accrete. In fact, the case for gas in globular clusters is much stronger now than it was in the 1970's. More detailed models have been made for how much gas should be retained (Pfahl & Rappaport 2001), and these models are in agreement with the two clusters where estimates have been made of the gas content by Freire and collaborators (2001) based on the correlations between pulsar dispersion measures (indicative of the amount of gas along the line of sight to the pulsar) and pulsars period derivatives (indicative of the direction in which the pulsar is being accelerated, and hence giving the de-projected distance from the center of the globular cluster along the line-of-sight).

Unfortunately, if one takes a conservative estimate of the mass accretion rates provided by this gas (see e.g Perna et al. 2003; Pellegrini 2005), most of the proposed globular cluster black holes would still not be detectable in the X-rays (Maccarone 2004). Some of them would, however, be detectable in the radio. A particularly interesting case is the example of G1, the largest globular cluster in M31. The claimed black hole in this globular cluster is about 20000 solar masses (Gebhardt, Rich & Ho 2002,2005). This is large enough that accretion of gas from the interstellar medium in the cluster might be expected to yield appreciable X-ray emission (Pooley & Rappaport 2006). X-ray emission is, in fact, seen from this cluster (Trudolyubov & Priedhorsky 2004). Furthermore, at that level of X-ray emission, one would expect appreciable radio emission from an intermediate mass black hole, but not from an accreting stellar mass black hole or an accreting neutron star, making it possible to break the degeneracy among the most likely mechanisms for the X-ray emission (Maccarone & Koerding 2006). Radio emission at the predicted flux level was detected by Ulvestad et al. (2007).

**Stellar black holes in globular clusters**

The focus of searches for black holes in globular clusters has generally been on searches for intermediate mass black holes. However, it has long been an open question about whether even stellar mass black holes could exist in globular clusters. With the advent of the Chandra and XMM observatories, it became possible to test this more carefully, as we finally have highly sensitive telescopes with sufficient angular resolution to allow firm associations between extragalactic X-ray binaries and globular clusters.

Chandra quickly observed X-ray sources brighter than the Eddington limit for a neutron star in several extragalactic globular clusters (Angelini, Loewenstein & Mushtozky 2001; Sarazin, Irwin & Bregman 2001), but these sources were not "smoking guns." Since extragalactic globular clusters have cores smaller than the Chandra point spread function, one could not be sure that these were single sources, rather than superpositions of several bright neutron stars; to prove this definitively with current instrumentation requires large amplitude variability (Kalogera, King & Rasio 2004).

Recently, we found such a source, in the galaxy NGC 4472 (Maccarone et al. 2007), a source at $4*10^{39}$ ergs/sec which varied by a factor of about 7 in count rate in a few hours. Furthermore, the globular cluster hosting this source has been spectroscopically identified as a globular cluster, removing the possibility that this object is a background active galactic nucleus. The X-ray spectrum of this object is extremely soft and disk blackbody fits give a characteristic radius of a few hundred kilometers, consistent with either a few hundred solar mass black hole, or with the super-Eddington disk models proposed for ultraluminous X-ray sources (Mukai et al. 2003; King & Pounds 2003), where radiation pressure does not stop accretion entirely, but rather limits it to the local Eddington limit, allowing for an increase in the actual maximum luminosity by the logarithm of the ratio of the actual mass accretion rate to the accretion rate required to produce the Eddington luminosity.

Many of this object's properties resemble the 1989 outburst of the Galactic X-ray binary V404 Cygni (compare e.g. with Oosterbroek et al. 1997 and Zycki et al. 1999) – the variability is almost exclusively at low energies, consistent with changes in the absorption column, probably from a disk wind. Furthermore, the source appears to have an accretion disk temperature much less than is usually observed for high luminosity black hole accretors, and an inner radius for the accretion disk considerably bigger than is usually observed. The disk wind would be most likely for a source accreting at a large fraction of the Eddington limit, so this object is most likely a stellar mass black hole. The inner disk apparent radius and temperature are also consistent with the expectations from a super-Eddington accretor.

The strong evidence for a black hole in a globular cluster thus re-opens the possibility that our own Galactic globular clusters may contain black holes. Obviously, we do not see any strongly accreting black holes in the Galaxy's cluster system, but the theoretical scenarios proposed by Kalogera et al (2004), for example, suggest that if such objects exist, they should be low duty cycle transients. Therefore, they may be hidden among the low luminosity X-ray sources which are highly abundant in the Galactic globular clusters. Quiescent black hole X-ray binaries would be quite difficult to distinguish from cataclysmic variable stars – the optical spectra of dwarf novae in quiescence show no strong differences from the optical spectra of black hole X-ray binaries in quiescence, with the possible exceptions of a slightly lower flux level of He II 4686 emission in black hole systems (although this line is also difficult to detect in dwarf novae – see e.g. Marsh et al. 1994). Black hole X-ray binaries might be expected to show larger ratios of optical to X-ray flux for a given orbital period, but only in 47 Tuc are there are a large number of objects with orbital periods, as well as good X-ray coverage (Edmonds et al. 2003). It should be noted that one object in that cluster, W21, does have an unusually large optical to X-ray flux ratio for a CV with its orbital period, but that that ratio is reasonable for a stellar mass black hole.

The smoking gun would most likely be to detect radio emission from the black hole candidate – while black holes in quiescence are weak radio sources, they have rather large ratios of radio-to-X-ray flux (e.g. Gallo et al. 2006). Unfortunately, the characteristic flux levels expected are about 10 uJy for relatively nearby globular clusters, and are reachable in reasonable exposure times only with Northern Hemisphere radio telescopes, while the most interesting globular clusters (i.e. the ones which are relatively nearby and have high stellar interaction rates) are primarily in the Southern hemisphere.

**Single White Dwarfs in Globular Clusters**

White dwarfs are the compact remnants of low-mass stars like the Sun. The vast majority of a cluster's original stellar population (and even of its "dead" population) would have belonged to this category, and thus white dwarfs should outnumber all other types of compact objects in globular clusters by far. Physically, white dwarfs are rather simple objects, since they possess no significant internal energy source. Rather like lumps of heated metal, they just gradually cool down as they radiate away whatever heat they retain from their previous lives. This simplicity is extremely useful. Most importantly, it means that the age of a globular cluster can be reliably inferred from the temperature of the coolest white dwarfs it contains (e.g. Hansen 1999). Such cluster ages are of great astronomical interest, since they place firm constraints on the era of star and galaxy formation in the universe.

White dwarfs are notoriously difficult to find in globular clusters, however. The problem is that their radii are about 100 times smaller than those of ordinary stars, which, other things being equal, makes them 10,000 times fainter. Moreover, if the goal is to estimate a cluster's age, the most interesting white dwarfs are the very coolest, and hence the very faintest. Locating such objects in a crowded cluster packed with normal stars is a serious challenge, and both ground-based (e.g. Richer 1978, Chan & Richer 1986, Ortolani & Rosino 1987, Richer & Fahlman 1988) and even early HST-based searches (e.g. Elson et al. 1995, Paresce, De Marchi & Romaniello 1995) struggled to uncover more than a handful of relatively bright white dwarf candidates in any given cluster.

Somewhat larger white dwarf samples were constructed by Richer et al. (1995) and Renzini et al. (1996), but it took several more years before ultra-deep HST observations of M4 yielded the first white dwarf-based age estimate of an ancient GC (12 Gyr; Richer et al. 2002, Hansen et al. 2002, Hansen et al. 2004). Since then, the latest addition to HST's arsenal of instruments – the *Advanced Camera for Surveys* (ACS) – has been used to obtain even deeper and shaper images of another cluster, NGC 6397 (Figure 2; Richer et al. 2006). Those observations have been turned into the most reliable and precise age estimate for any GC to date, 11.47 +/- 0.47 Gyr at 95% confidence (Hansen et al. 2007). It is worth putting these results into a cosmological perspective: the white dwarf-based ages for M4 and NGC 6397 imply that the star formation in these clusters took place at redshifts of $z \sim 3$. This coincides nicely with the epoch of maximum star formation that has been inferred (much more indirectly) from deep cosmological galaxy surveys (e.g. Madau et al. 1996,

Thompson et al. 2006).

**White Dwarfs in Binaries**

A large fraction of stars in the Milky Way are members of binary or higher-order multiple systems (although the long-held belief that *most* stars are in multiples has recently been challenged; Lada 2006). Since white dwarfs are the most common end-point of stellar evolution, their binary fraction in the Galactic field is also expected to be quite high, and this does appear to be borne out observationally (Holberg et al. 2002). However, this finding cannot be translated directly to white dwarfs in globular clusters, since dynamical interactions can both destroy wide binaries and create close ones (e.g. Hut et al. 1992). Nevertheless, both theoretical and empirical work suggests that, even in dense cluster cores, 5% - 50% of all stars are members of binary systems (Albrow et al. 2001; Ivanova et al. 2005; Sollima et al. 2007; Hurley, Aarseth & Shara 2007). Presumably, the white dwarf binary fraction in clusters should then be appreciable as well.

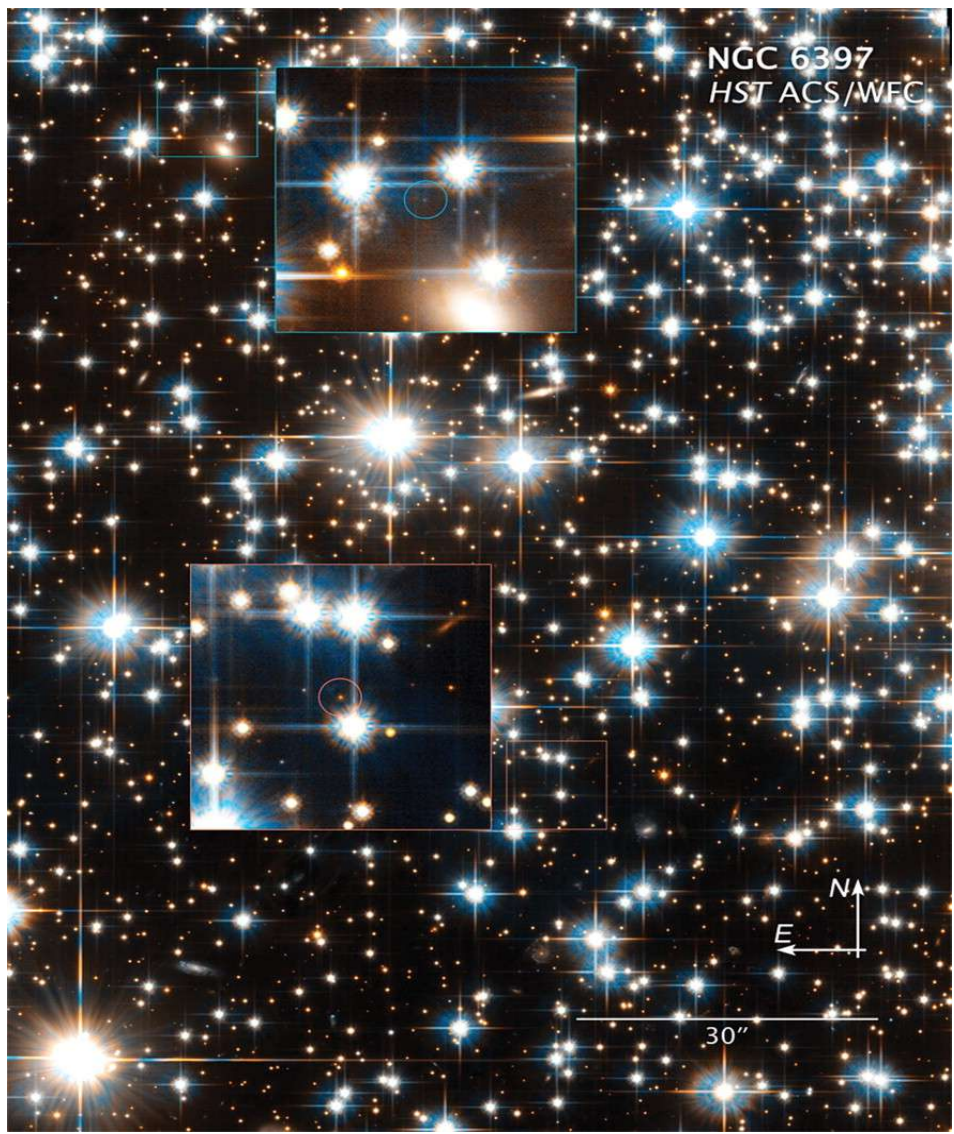

*Figure 2: An ultra-deep optical image of a field in the globular cluster NGC 6397, as obtained with*

*the Advanced Camera for Surveys onboard HST. The top inset shows a very low mass main sequence star in the cluster. The bottom inset shows one of the cool white dwarfs that were found, which were used to determine a precise age of 11.5 billion years for this cluster.*

Why should we care about white dwarf binaries in globular clusters? One reason is simply that the presence of binaries makes it more difficult to construct clean samples of single white dwarfs, such as those needed for the purpose of age-determination (Hurley & Shara 2003). However, the more exciting reason is that the broad class of "white dwarf binaries" includes a variety of fascinating and exotic systems with special importance, not just for the study of globular clusters, but for astrophysics more generally.

**Accreting White Dwarfs**

The first class of cluster white dwarf binaries to receive considerable attention were the so-called cataclysmic variables (CVs). These are the white dwarf counterparts to the low-mass X-ray binaries mentioned previously, i.e. they are close binary systems in which a white dwarf (rather than a neutron star or black hole) accretes material from nearby main sequence companion. The realization that globular clusters contained more than their fair share of X-ray bright LMXBs immediately led to the expectation that these clusters should also harbor large populations of CVs. After all, there are 100 to 1000 times more white dwarfs than neutron stars in any given cluster, so CVs might be expected to outnumber LMXBs by a similar margin (e.g. Bailyn 1991). This would make CVs extremely useful tracers of the dynamical processes that produce compact binaries in globular clusters.

The discovery of a significant population of "dim" X-ray sources in globular clusters (with $L_x < 10^{34.5}$ erg/s) seemed to broadly support the idea that quite large CV populations might reside there (Hertz & Grindlay 1983ab; Hertz & Wood 1985). However, early X-ray surveys were hampered significantly by the limited sensitivity and angular resolution available at the time, so fairly drastic extrapolations were necessary to convert the observed X-ray fluxes into a corresponding number of CVs.

In order to conduct a more reliable census, researchers therefore turned to other wavelengths and discovery methods. In the Galactic field, CVs are generally found to be (i) exceptionally blue, (ii) highly variable, and (iii) strong emission line sources (e.g. Warner 1995). Each of these distinguishing characteristics was used as the basis for CV searches in globular clusters (e.g. (i): Paresce, de Marchi & Ferraro 1992; Ferraro et al. 1997; (ii): Shara, Bergeron & Moffat 1994; Shara et al. 1996; (iii): Cool et al. 1995; Grindlay et al. 1995). However, even though all of these methods discovered a few systems, the large CV populations that had been expected -- tens to hundreds of systems in each cluster – failed to turn up.

In tandem with these observational developments, the theoretical predictions for CVs in clusters were also being refined. Numerical models for CV formation and evolution in a dense cluster environments were used to predict that a single massive, dense cluster like 47 Tuc might contain

~200 CVs formed by dynamical encounters between white dwarfs and single stars (Di Stefano & Rappaport 1994). Moreover, at least two other channels – CV formation directly from primordial cluster binaries (Davies 1997) and via 3-body encounters between single stars and binaries (Davies & Benz 1995) – seemed theoretically capable of producing comparable numbers. It should be noted that all of these theoretical results were rather uncertain, partly because they involved poorly understood physical processes (e.g. the orbital shrinkage of pre-CVs during the so-called common envelope phase, or the response of a main sequence star to being captured by a white dwarf), and partly because significant approximations had to be made to keep the computations tractable. Even today, a full N-body simulation of a cluster like 47 Tuc, with a detailed accounting for stellar and binary evolution, is not quite within reach. However, the latest set of theoretical simulations (Ivanova et al. 2006) still predicts that ~200 CVs in total should reside in a cluster like 47 Tuc, a result that remains roughly consistent with the earlier estimates.

Given all this, the situation regarding CVs in globular clusters was rather unclear at the end of the 20$^{th}$ century. Did clusters harbor the large expected CV populations or not? If they did, why had we failed to find them? Were cluster CVs somehow different from their counterparts the Galactic field? And if large populations of CV did not exist, did this mean our understanding of the dynamical processes that create and destroy binaries in clusters was wrong?

Fortunately, the first years of the 21$^{st}$ century have already seen several significant developments that are beginning to provide answers to these questions. The most significant progress resulted from the ability to carry out much deeper and sharper X-ray surveys of globular clusters, especially with the *Chandra X-ray Observatory*. The very first such survey – of 47 Tuc – immediately revealed the existence of over a hundred, mostly very faint ($10^{30} <\sim L_x <\sim 10^{31}$ erg/s), X-ray sources in this cluster (Grindlay et al. 2001; an extension of this survey to even greater depth was reported by Heinke et al. 2005). Based mainly on their X-ray colours, approximately 30 sources were considered to be likely CVs, and this expectation was more or less borne out by subsequent optical follow-up studies with HST (Edmonds et al. 2001ab).

Since then, other clusters have also been studied with Chandra (e.g. Pooley et al. 2002; Heinke et al. 2003), as well as with ESA's *X-ray Multiple Mirror Telescope* (XMM; e.g. Gendre, Barret & Webb 2003; Webb, Wheatley & Barret 2006). A global analysis of the faint X-ray source populations that have been uncovered by such studies shows that, once the effect of cluster mass is taken into account, the number of faint X-ray sources in a given cluster depends strongly on the frequency with which the stars in that cluster stars undergo close encounters (Pooley & Hut 2006). This result strongly suggests that the CVs that are now being found in globular clusters are mainly formed by dynamical processes, rather than directly from primordial binaries.

Another useful development in the hunt for CVs in globular clusters was the deployment of first the *Space Telescope Imaging Spectrograph* (STIS) and later ACS onboard HST. Amongst their many other capabilities, these instruments allowed us for the first time to take deep, high-resolution images of globular clusters at far-ultraviolet wavelengths (e.g. Brown et al. 2001; Knigge et al. 2002). This waveband offers two related advantages to CV searches: first, CVs are very blue objects (as already noted above), and many radiate most of their energy in the far-ultraviolet (van Teeseling & Verbunt 1994; Verbunt et al. 1997). Second, most of the other stars in a globular cluster produce negligible far-ultraviolet flux. The second point matters, because it is the extreme crowding that makes it so difficult to optically identify CVs in dense cluster cores. At optical wavelengths, the few CVs we are searching for are faint and easy to miss amongst the sea of often much brighter "normal" stars. At far-ultraviolet wavelengths the situation is reversed. Here, CVs far outshine other cluster members, making it much easier to identify and study them. The situation is illustrated in Figure 3, which shows the same field in the core of 47 Tuc, imaged in both optical and far-ultraviolet wavebands. The far-ultraviolet image is far less croded, since most of the normal stars in the optical image produce almost no far-ultraviolet radiation and thus simply disappear out of sight. The brightest object in the far-ultraviolet image is a CV (Knigge et al. 2002, 2003). Several clusters have by now been imaged in the far-ultraviolet and searched for CV candidates (e.g. Knigge et al. 2002; Dieball et al. 2005; Dieball et al. 2007). In all cases, 10-100 objects with the colours expected for CVs were found, although most still need to be confirmed as *accreting* objects (as opposed to, for example, detached white dwarf – main sequence binaries).

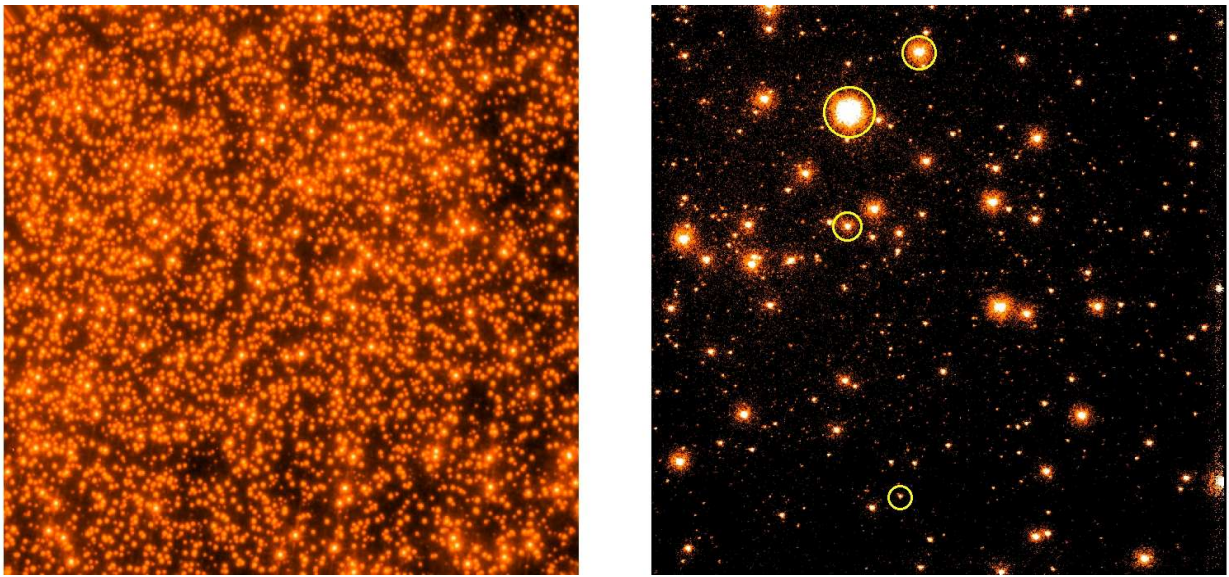

*Figure 3: The core of the dense globular cluster 47 Tuc as it appears in optical (left panel) and far-ultraviolet (right panel) images obtained with HST. Note that the far-ultraviolet image is much less crowded, since the ordinary stars that dominate at optical wavelengths are too cool to generate a significant amount of far-ultraviolet light. As a result, unusually hot stars – such as cataclysmic variables – can be much more readily identified in the right hand image. The four objects circled in yellow – which include the brightest far-ultraviolet source in the cluster – are all cataclysmics.*

So has the "CV problem" in globular clusters been solved? Not quite. It is certainly reassuring that X-ray and far-ultraviolet surveys are finally turning up sizable populations of CV candidates, and that clusters with higher dynamical encounter rates appear to produce larger populations. However, the observed populations (~tens) are still considerably smaller than the predicted ones (~hundreds). It remains unclear whether this simply reflects the difficulty of finding the very faintest CVs -- which are theoretically expected to dominate the overall CV populations (e.g. Kolb

1993; Di Stefano & Rappaport 1994) – or if there is a genuine dearth of CVs in clusters compared to theoretical predictions.

There is also some evidence that cluster CVs may be somewhat different from their cousins in the Galactic field. For example, it has proven difficult to simultaneously reconcile the X-ray and optical properties of the 47 Tuc CVs with those of known field CVs (Edmonds et al. 2003b). Similarly, based on the low numbers of cluster CVs found by variability searches (e.g. Shara, Bergeron & Moffat 1994; Shara et al. 1996; Shara, Hinkley & Zurek 2005; Kaluzny et al. 2005; Bond et al. 2005), it has been suggested that the variability properties of cluster CVs may be different, possibly because they preferentially contain magnetic white dwarfs (e.g. Grindlay 1999; Dobrotka, Lasota & Menou 2006).

It is too early to tell if these are genuine problems, mainly because existing CV samples suffer from serious selection effects. For example, it is well-known that it is the faintest CVs (which should dominate the overall population) are also the least variable. Thus the low numbers of CVs turned up by variability searches may simply reflect the expected dominance of faint, rarely erupting systems in the overall population. Moreover, cluster and field CV samples probably suffer from quite *different* selection effects. Most of the former have so far been found via X-ray searches, whereas this is true for only a minority of the latter (Gaensicke 2005) . Thus great care has to be taken when comparing field and cluster samples. Finally, even if there are really fewer CVs in clusters than predicted, this need not necessarily mean that our understanding of binary formation via cluster dynamics is at fault.

An alternative possibility is that we do not understand the binary evolution of CVs. The same CV evolution scenarios that are being used to predict CV numbers in clusters have long predicted the existence of a large population of faint field CVs... which has also remained rather elusive. Even though some CVs belonging to this predicted population are finally turning up int the Galactic field (e.g. Littlefair et al. 2006), there is strong evidence that selection effects cannot be (solely) to blame for the small numbers detected so far (e.g. Pretorius, Knigge & Kolb 2007). If there really is something wrong with CV evolution theory, predictions for cluster CVs will certainly be affected as well.

**Double White Dwarfs: Ticking Time Bombs and Gravitational Wave Sources**

A second class of white dwarf binaries that deserve special attention are the *double* white dwarfs. A typical white dwarf is only about as big as the Earth, despite having a mass comparable to that of the Sun. This compactness means that white dwarf - white dwarf binaries can be extremely tight, with orbital periods as short as minutes.

Double white dwarf binaries can be formed in clusters via the same channels as white dwarf - main sequence binaries and binaries involving neutron stars, although the relative importance of the various channels is expected to be different in each case (Shara & Hurley 2002; Ivanova et al. 2004). There are also double white dwarf analogues of CVs, i.e. systems in which one white dwarf accretes material from another. Such systems are known as AM CVn stars, and about twenty of these have so far been identified in the Galactic field (e.g. Nelemans 2005). No AM CVn star has yet been found in a globular cluster, although such systems are predicted to exist.

The binary evolution of double white dwarfs – both detached systems and AM CVns – is mainly driven by gravitational radiation. As the two massive stars orbit each other, their motion creates ripples in space-time that spread out like water waves. These gravitational waves carry both energy and angular momentum, which are thus drained from the binary itself. As a result, the binary shrinks, and its orbital period decreases. In most binary systems, gravitational radiation produces only very slow orbital shrinkage, so that other mechanisms usually dominate the binary evolution. However, in a compact double white dwarf binary system, gravitational radiation can have much more dramatic effects, and the drain of energy and angular momentum from the orbit can be very fast. This has several important consequences.

First, it means that the AM CVn systems can have high accretion rates and luminosities, since these are roughly proportional to the gravitational-radiation-induced rate of period change. Thus AM CVns can actually be rather bright.

Second, short-period ($P_{orb}$ <~ 30 min) double white dwarfs in globular clusters may be detectable as gravitational wave sources by LISA, the *Laser Interferometer Space Antenna*, which is due to launch in 2015 (Benacquista, Portegies Zwart & Rasio 2001; Ivanova et al. 2006). In fact, white dwarf – white dwarf binaries in the Galactic field are expected to be so common that they will act as a significant source of foreground noise for LISA (e.g. Ruiter et al. 2007). Double white dwarfs in globular clusters should be detectable above this noise threshold, however, since the vast majority of field binaries reside in the plane of the Galactic disk. Most globular clusters are located well away from this plane, in areas of the sky where the noise level is considerably lower. Interestingly, dynamical interactions in clusters may also produce a significant number of *eccentric* double white dwarfs, which have no counterparts in the Galactic field at all (Willems et al. 2007). These, too, may be detectable with LISA.

Third, the orbital shrinkage caused by the emission of gravitational radiation can be so fast that the binary system may merge on a time-scale shorter than the age of the Universe. For example, a double white dwarf system with $P_{orb}$ ~ 30 min (such as those detectable by LISA) is expected to merge on a time-scale of a few million years (e.g. Benacquista, Portegies Zwart & Rasio 2001). The outcome of such a merger depends primarily on the combined mass of the binary components. For system masses in excess of about 1.4 solar masses (the so-called Chandrasekhar limit, which represents the maximum mass of a single white dwarf ), it is generally expected that the merger will result in a type Ia supernova explosion. It has therefore been suggested that the dynamical production of double white dwarfs could make globular clusters efficient "Type Ia Supernova Factories" (Shara & Hurley 2002). At least in elliptical galaxies, the fractional contribution of cluster sources to the overall SN Ia rates could be very significant (Ivanova et al. 2004).

**Summary**

The past decade has seen a dramatic growth in our understanding of the compact object populations of globular clusters, almost all driven by new facilities – the Hubble Space Telescope, as well as the solar Blind Camera for white dwarf work; the Green Bank Telescope and especially the Pulsar Spigot for radio pulsar work; and the Chandra and XMM-Newton observatories for searches for accreting black holes and neutron stars. The near future should be bright for searches for faint radio sources in clusters, with e-Merlin due to give very high sensitivity while still giving sufficient angular resolution to avoid source confusion problems and to allow for easy identification of optical counterparts.